\documentclass[final,5p,times,twocolumn,authoryear]{elsarticle}

\usepackage{amssymb}
\usepackage{lipsum}
\usepackage{physics}
\usepackage{amsmath}
\usepackage{amsfonts,amssymb}
\usepackage{graphicx,graphics}
\usepackage{hyperref}
\usepackage{color}
\usepackage{mathrsfs}
\usepackage{bbold}
\usepackage[normalem]{ulem}

\journal{Physics Letters B}

\begin{document}

\begin{frontmatter}

\title{Quantum detection of boundary conditions}

%\author[first]{Jefferson Mon\'c\~ao da Silva}    
%\affiliation[first]{organization={Departamento de Matem\'atica Aplicada, Universidade Estadual de Campinas}, Departamento de Matem\'atica Aplicada          
%            postcode={13083-859}, 
 %           city={Campinas},
  %          state={S\~ao Paulo},
   %         country={Brazil}}
            
%\author[second]{Jo\~ao Paulo M. Pitelli}
 %           \affiliation[second]{organization={Departamento de Matem\'atica Aplicada, Universidade Estadual de Campinas}, Departamento de Matem\'atica Aplicada          
 %           postcode={13083-859}, 
  %          city={Campinas},
   %         state={S\~ao Paulo},
    %        country={Brazil}}
\author[uos]{Jefferson Mon\c{c}\~ao da Silva}
\ead{j218456@dac.unicamp.br}
\author[uos]{Jo\~ao Paulo M. Pitelli}
\ead{pitelli@unicamp.br}
\address[uos]{Departamento de Matem\'atica Aplicada, Universidade Estadual de Campinas, 13083-859 Campinas, S\~ao Paulo, Brazil}

\begin{abstract}
The quantum detection of spacetime  conicity  was investigated in reference [Phys. Lett. B 820, 136482 (2021)] through the analysis of the response function of an Unruh-deWitt detector coupled to a massless scalar field. In particular, it was shown  that the detector can discern the presence of the deficit angle of a region of spacetime  even when it is placed on a  flat region with no conical deficit and (due to the rapid switching on and off) is  classically isolated from the  conical region. The scalar field was supposed to be continuous at the three dimensional timelike hypersurface connecting both regions and it was argued that the non-trivial effects on the response function were only due to the deficit angle of the outer conical region. In this paper, we study the response function of the Unruh-deWitt detector on the same classical framework, but exploring the effects  of the choice of boundary conditions for the scalar field at the connecting hypersurface. These boundary conditions are extracted from the two-interval Sturm-Liouville theory and  each  one gives rise to  a physically inequivalent sensible dynamics for the quantum field. We show that they play an important role on the response function of the detector, which effectively ``feels'' the combined effect of conicity + boundary condition.
\end{abstract}


\end{frontmatter}

\section{Introduction}
\label{introduction}
It is well known that Klein-Gordon equation for the field $\hat{\Psi}$ (as well as any second-order hyperbolic equation) in globally hyperbolic spacetimes is well-posed~\citep{wald}. This means that the time evolution $\hat{\Psi}(t)$ is uniquely determined by (and depending continuously on) the initial data $\hat{\Psi}(0)$ and $n^{\mu}\nabla_\mu\hat{\Psi}(0)$ on a Cauchy surface  with normal vector $n^\mu$. This is not the case in non-globally-hyperbolic spacetimes, where there is no such Cauchy surface and initial data on any achronal hypersurface are not enough to determine uniquely the evolution of the field. 

 In  spacetimes with timelike boundaries, the domain of dependence of any closed achronal hypersurface covers only part of the spacetime - the rest of it being influenced by the boundary.  Nevertheless, it is still possible to define ``sensible dynamics'' for the field $\hat{\Psi}$ in  static non-globally-hyperbolic spacetimes given initial data on a spacelike hypersurface  $\Sigma$. The prescription consists on the choice of a (possibly non unique) positive self-adjoint extension of the spatial part of the wave operator~\citep{wald_1980}. The main aspects of these sensible dynamics are: i) the agreement with the solution of the Klein-Gordon equation inside the domain of dependence $D(\Sigma)$ (with initial data in $C_0^\infty(\Sigma)$), ii) compatibility with causality and iii) the existence of a conserved energy functional~\citep{ishibashi}.  In general, there will be an infinite number of inequivalent sensible dynamics, corresponding to different positive self-adjoint extensions of the spatial part of the wave operator. To each positive self-adjoint extension, there corresponds  a  boundary condition at the boundary of spacetime (these boundary conditions will, from now on, be called self-adjoint boundary conditions).
 
In a static spacetime with a single timelike boundary, the spatial part of the wave equation turns out to be a one-interval Sturm-Liouville. In adapted coordinates (where the boundary is ``located'' at the spatial coordinate $x=a$), the wave equation becomes (after separating variables in the form $\hat{\Psi}=e^{-i\omega t}\Xi(x^i)$)
\begin{equation}
A\Xi=\omega^2\Xi,\,\,\,\,x>a,
\end{equation}
where $A$ denotes the spatial part of the wave equation. Clearly, a boundary condition at $x=a$ is necessary (the other boundary would be at $b=\infty$, where square-integrability usually works as a natural boundary condition). In the same way, consider a static spacetime between two timelike boundaries at $x=a$ and $x=b$.  The spatial part of the wave equation is (once again) a  one-interval Sturm-Liouville of the form
\begin{equation}
A\Xi=\omega^2\Xi,\,\,\,\,a<x<b,
\end{equation}
with  boundary conditions imposed at $x=a$ and $x=b$.

In this paper, we consider, instead, two spacetime regions separated by a timelike boundary at $x=x_0$. Within this configuration, the spatial part of the wave equation should be split into two Sturm-Liouville problems, one for each separated regions. We arrive at
\begin{equation}\left\{
\begin{array}{lll}
A_1\Xi=-\omega^2\Xi, & a<x<x_0, & \textrm{(first region)},\\
A_2\Xi=-\omega^2\Xi, & x_0<x<b, & \textrm{(second region)},
\end{array}\right.
\end{equation}
with boundary conditions at $x=x_0^-$, $x=x_0^+$, $x=a$ and $x=b$. In general, the smooth condition $\hat{\Psi}(x=x_0^-)=\hat{\Psi}(x=x_0^+)$ and $\hat{\Psi}'(x=x_0^-)=\hat{\Psi}'(x=x_0^+)$ is only one among an  infinite number of self-adjoint boundary conditions - all of them giving rise to physically sensible dynamics for the field. The mathematical theory behind these self-adjoint boundary conditions is called two-interval Sturm-Liouville problem (see, for instance, Refs.~\citep{everitt, zettl, wang}). 

{
In the present work, the separation between the first and second regions is defined by a timelike singular shell that splits the spacetime into two regions with distinct conicities (see Sec.~\ref{secii}). This shell carries an energy density due to the violation of Israel's second junction condition by the metric. Because the derivative of the metric is discontinuous, the Klein--Gordon equation includes non-distributional terms. As discussed in Ref.~\cite{pitelliwire} in a much simpler context, the presence of non-integrable terms makes physical observables fully dependent on how the singularity is regularized. 

Such regularizations can be effectively modeled through boundary conditions imposed at the singularity (with the smooth condition being only of the possibilities). This reflects our limited understanding of how singularities affect the surrounding spacetime. In practice, we encode this ignorance via self-adjoint boundary conditions - here the smooth boundary condition implies no interaction between the shell and the field. It is worth emphasizing that this situation differs sharply from cases involving only integrable potentials, where physical results are independent of the chosen regularization (see Ref.~\cite{andrews}).
}

In order to simplify many of our discussions, as well as to compare our results with some known  results from the literature, we will consider a flat spacetime (with no conical deficit) as the inner region, which is connected (via a timelike shell) to a conical outer region. This framework was used by Cong et al. in Ref.~\citep{cong}.  There,  an Unruh-deWitt detector was placed on the inner region and the interaction time with the quantum field was rapid enough so that no signal could travel from the detector to the shell and back to the detector. It was shown that, due to the non-local character of the vacuum state, the fluctuations of the field around the detector convey information about the conicity of the outer region. This can be seen by the non-trivial behavior of the response function (when compared to the response function of a detector in Minkowski spacetime).

The main goal of this paper is to explore the non-locality of  quantum phenomena in quantum field theory in static and non-globally-hyperbolic curved spaces. In particular, we discuss how changes in the  boundary conditions in a distant region of space can affect the observer in the quantum  scenario. In the proposed framework, the Unruh-DeWitt detector is situated within a flat region $\mathcal{V^-}$,  bounded by a cylindrical shell $S$, outside of which there is an external region $\mathcal{V^+}$ with a conical deficit. We are interested in investigating the Unruh-DeWitt detector's sensitivity to different boundary conditions on the junction surface \( S \). We, therefore, substitute the usual smooth condition considered in~\citep{cong}
\begin{equation}
\begin{array}{r}
    [\hat{\Psi}(x)] = 0,  \\
    {[l^{\mu}\nabla_{\mu}\hat{\Psi}(x)]} = 0,
\end{array} 
\label{smooth}
\end{equation}
where \( l^{\mu} \) is normal to $S$ and $[A(x)]\equiv\left.A( \mathcal{V}^+)\right|_{S}-\left.A( \mathcal{V}^-)\right|_{S}$, by more general self-adjoint boundary condition obtained by the application of the two-interval Sturm-Liouville problem.

\section{The flat/conical space-time}
\label{secii}
 Since we are considering non-globally hyperbolic and static spacetimes, let us discuss a few aspects of them. A spacetime $\mathcal{M}$ is called stationary if it has a time-like Killing field $t^\alpha$. In this case, in the coordinate system $(t,x^i)$ where $t^{\alpha}\equiv \delta^{\alpha}_t$, the metric does not depend on the coordinate $t$. A stationary spacetime $\mathcal{M}$ is also called static if the metric is invariant under time reversal, $t \rightarrow -t$.  In a static spacetime, the metric takes the following form
\begin{equation}
ds^2=-V^2dt^2+h_{ij}dx^idx^j, \label{metric nglobal}
\end{equation}
where $V(x^i)$ and $h_{ij}(x^i)$ are independent of $t$ and  $V^{2}= - t_\alpha t^\alpha$. Our focus lies on spacetimes $\mathcal{M}$ having a hypersurface $\Sigma$ orthogonal to $t^\alpha$, with  the orbits of $t^\alpha$ intersecting $\Sigma$ only once. Considering the metric form given by Eq. (\ref{metric nglobal}), the volume element is defined as $V^{-1}d\Sigma$ where $d\Sigma$ is the volume element of the hypersurface $\Sigma$, and we define $\Sigma_t$ as the ``time translation'' of $\Sigma$ by the flow $t^{\alpha}$. 

The Klein-Gordon equation for the massless scalar field can be written in the form
\begin{equation}
    \frac{\partial^2\hat{\Psi}}{\partial t^2}=-A\hat{\Psi},
    \label{KG Eq}
\end{equation}
where $A=-VD^i(VD_i)$ and $D_i$ is the covariant derivative on $\Sigma_t$. With the ``initial'' domain $D(A)=C_0^{\infty}(\Sigma)$, $A$ is not self-adjoint. However, there will be (at least one) positive self-adjoint extensions for the operator $(A,\mathcal{D}(A))$~\citep{simon}. These positive self-adjoint extensions are denoted by $(A_\beta,\mathcal{D}(A_\beta))$, where $\beta$ denotes a (possibly infinite many) parameter. The ``sensible'' physical evolution is defined as the solution of the equation~\citep{wald_1980}
\begin{equation}
    \frac{d^2\hat{\Psi}}{dt^2}=-A_\beta\hat{\Psi},
\end{equation}
with $d/dt$ being the derivative on the Hilbert space $L^2(\Sigma,V^{-1}d\Sigma)$, i.e., 
\begin{equation}
    \hat{\Psi}(t)=\cos{(\sqrt{A_\beta}t)}\hat{\Psi}(0)+\frac{1}{\sqrt{A_\beta}}\sin{(\sqrt{A_\beta}t)}\dot{\hat{\Psi}}(0). 
\end{equation}

The spacetime  under consideration comprises two regions, denoted as $\mathcal{V}^-$ and $\mathcal{V}^+$, each endowed with its respective coordinate system: $(t, z, \rho_-, \phi)$ and $(t, z, \rho_+, \phi)$, where $0 \leq \rho_- \leq R_1$ and $cR_1 \leq \rho_+ \leq cR_2$ with $R_1 \leq R_2$. The metric\footnote{Notice that the metric in the region $\mathcal{V}^+$ is conical and,  therefore, should exhibit a reduced angular interval. However, due to the choice of coordinates $\rho_-$ and $\rho_+$, the angular interval runs from $\phi=0$ to $\phi=2\pi$ on both regions. In coordinates such that $0<\rho<R_2$,  we have  that  both $\rho_-=R_1$ and $\rho_+=c R_1$ correspond the same point $\rho=R_1$.} in each region is given by~\citep{cong}
\begin{equation}
    \begin{array}{ll}
         ds_-^2  = -dt^2 + dz^2 + d\rho_-^2 + \rho_-^2 d\phi^2, & \textrm{on}\,\,\mathcal{V}^- \\
         ds_+^2  = -dt^2 + dz^2 + d\rho_+^2 + \frac{\rho_+^2}{c^2} d\phi^2, & \textrm{on}\,\,\mathcal{V}^+
    \end{array}
\end{equation}
with $c>1$. This spacetime is endowed with a singular hypersurface $S$, where a boundary condition  for the field $\hat{\Psi}$ is required (following the prescription of Ref.~\citep{wald_1980}). Notice that $S$ is a singular surface having  a certain energy density. This can be confirmed through the violation of Israel second junction condition.

In principle, $R_2$ should be infinity to avoid boundary effects. However, a finite $R_2$ simplifies many of our calculations. In this way, we choose (as in Ref.~\citep{cong}) Dirichlet boundary condition for the field at $\rho_+=cR_2$, isolating the field $\hat{\Psi}$. When necessary, we can take $R_2\to\infty$ to simulate an infinite conical region.

\section{Solution for the massless scalar field}

Klein-Gordon equation in this spacetime becomes 
\begin{equation}
\frac{\partial^2 \hat{\Psi}}{\partial t^2}=-A_\pm \hat{\Psi}, 
\label{KG with A}
\end{equation}
with  $A_\pm = -\vec{\nabla}_\pm^2$, where $\vec{\nabla}_\pm^2$ is the Laplace-Beltrami operator on $\mathcal{V}^{\pm}$ given by
\begin{equation}\begin{array}{ll}
\vec{\nabla}_-^2=\frac{1}{\rho_-}\partial_{\rho_-}(\rho_- \partial_{\rho_-})+\frac{1}{\rho_-^2}\partial^2_\phi+\partial^2_z,\\
\vec{\nabla}_+^2=\frac{1}{\rho_+}\partial_{\rho_+}(\rho_+ \partial_{\rho_+})+\frac{c^2}{\rho_+^2}\partial^2_\phi+\partial^2_z.
\end{array}
\end{equation}

Since the spacetime is static and given the cylindrical symmetry, we can express the field $\hat{\Psi}(x)$ as 
\begin{equation}
    \hat{\Psi}_{kmq}(x)=N_{kmq}e^{-i\omega t}e^{ikz}e^{im\phi}\psi_{mq}(\rho_{\pm}), \label{modos decoposicao}
\end{equation}
where $\omega^2 =k^2+q^2$,  $N_{kmq}$ is a normalization constant, $m\in\mathbb{Z}$  and $\psi_{mq}$ satisfies an independent Bessel equation in each separate region, i.e.,
\begin{equation}
    \left\{\begin{array}{ll}
         -\left(\rho_-\psi'_{mq}(\rho_-)\right)'+\frac{m^2}{\rho_-}\psi_{mq}(\rho_-)= \rho_- q^2 \psi_{mq}(\rho_-), \\
         -\left(\frac{\rho_+}{c}\psi'_{mq}(\rho_+)\right)'+\frac{m^2 c}{\rho_+}\psi_{mq}(\rho_+)= \frac{\rho_+q^2}{c}\psi_{mq}(\rho_+).
    \end{array}\right. \label{spl radial}
\end{equation}
This is a two-interval Sturm-Liouville equation, with weight functions $\rho_-$ and $\rho_+/c$ respectively. The solution is expressed in terms of Bessel functions of the first and second kinds $J$ and  $Y$, respectively,  and  is  given by
\begin{equation}
  \left\{\begin{array}{ll}
    \psi_{mq}(\rho_-)=a_1 J_{m}(q\rho_-)+a_2 Y_{m}(q \rho_-), & 0<\rho_-<R_1\\

    \psi_{mq}(\rho_+)=c_1 J_{mc}(q\rho_+)+c_2 Y_{mc}(q \rho_+), & 0<\rho_+<c R_2.
    \end{array}\right.
    \label{bessels}
\end{equation}

The regularity condition of the field $\hat{\Psi}$  at $\rho_-=0$ eliminates the constant $a_2$ since  the function $Y_{m}$ diverges at this point. We then require $a_2=0$ so that
\begin{equation}
    \psi_{mq}(\rho_-)=J_{m}(q \rho_-),
\end{equation}
with  $a_1$ absorbed by the normalization constant $N_{kmq}$  (in other words, we say that $\rho_-=0$ is in the limit point case). However, to fully determine the scalar field mode, it is still necessary to choose a boundary condition at the junction surface $S$. But what are the correct (physical) boundary  conditions?

\section{Self-adjoint boundary conditions}
\label{Sec. boundary condition}

Let $k \geq 2$ be an integer. The multi-interval Sturm-Liouville problem in $k$ intervals consists in a system of $k$ differential equations
\begin{equation}
    -(p_r y')' + q_r y = \lambda \omega_r y \ \textrm{in} \ (a_r,b_r), \ r=1,\dots,k,
\end{equation}
with boundary conditions at the endpoints of each interval. Here, $\lambda \in \mathbb{C}$ is the fixed eigenvalue and $-\infty < a_r<b_r < \infty$ and~\citep{wang}
  \begin{equation}
       p_r^{-1}, q_r, \omega_r \in L(J_r,\mathbb{R}). 
   \end{equation}
Let us focus in the case $k=2$. In our case, $J_1 = (0, R_1)$  and $M_1$ is the following differential operator on $J_1$ with weight function $\omega_1 = \rho_-$,
\begin{equation}
    M_1 \psi_{mq}(\rho_-) = -\left(\rho_-\psi'_{mq}(\rho_-)\right)'+\frac{m^2}{\rho_-}\psi_{mq}(\rho_-).
\end{equation}
Similarly, $J_2 = (cR_1,cR_2)$ and  $M_2$ is the following differential operator in $J_2$  with weight function $\omega_2 = \rho_+/c$,
\begin{equation}
    M_2 \psi_{mq}(\rho_+) = -\left(\frac{\rho_+}{c}\psi'_{mq}(\rho_+)\right)'+\frac{m^2c}{\rho_+}\psi_{mq}(\rho_+).
\end{equation}

Our objective is to find self-adjoint boundary conditions in each interval $J_r$, so that the differential operator defined in $H = H_1 + H_2$, with $H_r = L^2(J_r,\omega_r)$, is self-adjoint. 
%can be obtained by the direct sum of self-adjoint operators in $H_1$ and $H_2$, these alone are insufficient to encompass all self-adjoint operators for the Sturm-Liouville problem in two intervals. Indeed, there exist many other operators involving interactions between the two intervals.

In general, elements of $H$ are represented as $\textbf{f} = (f_1,f_2)$ with $f_1 \in H_1$ and $f_2 \in H_2$, and the inner product in $H$ is defined by $(\textbf{f},\textbf{g})=(f_1,g_1)_1+(f_2,g_2)_2$, where $(\cdot,\cdot)_r$ denotes the usual inner product in $H_r$:
\begin{equation}
    (f,g)_r = \int_{J_r} f_r \bar{g}_r \omega_r.
\end{equation}

To characterize the self-adjoint solutions of this Sturm-Liouville problem, it is necessary to classify the boundary points in each interval. Note that in our specific problem, there are four boundary points. They are: $a_1=0, b_1=R_1, a_2 =cR_1$, and $b_2=cR_2$.
The point $a_1 =0$ is a limit point since the Neumann function, $Y_m(q \rho_-)$, diverges at $\rho_- =0$, and the other three endpoints points are regular. Additionally, although the points $\rho_-=R_1$ and $\rho_+=cR_1$ indicate the same spatial position at the junction surface $S_1$, they are considered as boundary points of distinct intervals.

In this case (with $a_1$ being a limit point), let $Y_i(x)\equiv \genfrac(){0pt}{0}{y_i(x)}{p_i y_i'(x)}$, $i=1,2$. Following Ref.~\citep{wang}, the self-adjoint boundary conditions are  described by the following equation
\begin{equation}
    A Y_1(b_1)+B Y_2(a_2)+C Y_2(b_2)=0,
    \label{eq matrix}
\end{equation}
where $A=(a_{ij})$, $B=(b_{ij})$, and $C=(c_{ij})$ are $3 \times 2$ matrices that satisfy the rank condition:
\begin{equation}
    \text{rank}(A|B|C)=3,
    \label{rank}
\end{equation}
and the self-adjoint condition:
\begin{equation}
    (a_{j1}\bar{a}_{k2}-a_{j2}\bar{a}_{k1})-(b_{j1}\bar{b}_{k2}-b_{j2}\bar{b}_{k1})+(c_{j1}\bar{c}_{k2}-c_{j2}\bar{c}_{k1})=0,
    \label{sa condition}
\end{equation}
where $a_{ij}, b_{ij}$, $c_{ij} \in \mathbb{C}$ and $j,k=1,2,3$.

As discussed earlier, we choose Dirichlet boundary condition at $\rho_+=cR_2$ so that there will be no exchange of energy-momentum of the field with the region $\rho_+>cR_2$ and the system can be considered isolated (as $R_2\to\infty$, we recover an asymptotically conical spacetime). Therefore, with this choice of boundary condition for the endpoint $b_2$, the  $C$ matrix  in Eq.~(\ref{eq matrix}) has the form
\begin{equation}
C = \begin{pmatrix}
    0 & 0 \\ 0 & 0 \\ 1 & 0
\end{pmatrix}
\end{equation}
while $A$ and $B$ becomes
\begin{equation}
A = \begin{pmatrix}
    a_{11} & a_{12} \\ a_{21} & a_{22} \\ 0 & 0
\end{pmatrix} \ \ {\text and} \ \ B = \begin{pmatrix}
    b_{11} & b_{12} \\ b_{21} & b_{22} \\ 0 & 0
\end{pmatrix}.
\end{equation}
In fact, with this definition of matrices $A$, $B$, and $C$, the third equation of (\ref{eq matrix}) is written as (in the case of interest):
\begin{equation}
\psi_{mq}(\rho_+=cR_2)=0.
\end{equation}

It can be shown (see Ref.~\citep{zettl}), that the boundary condition connecting the surface points $a_2$ and $b_1$ through Eq.~(\ref{eq matrix}) becomes [taking into account Eqs.~(\ref{rank}) and (\ref{sa condition})]
\begin{equation}
    Y(a_2)=e^{i\gamma}K \ Y(b_1)
    \label{bc field}
\end{equation}
with $- \pi < \gamma \leq \pi$ and $K\in SL_2(\mathbb{R})$ such that $det(K)=1$. 

In our case, we have
$Y(\rho)\equiv \genfrac(){0pt}{0}{\psi_{mq}(\rho)}{p_\pm \psi_{mq}'(\rho)}$ (here the notation $p_+ = \rho_+/c$ and $p_-=\rho_-$ is used) so that 
\begin{equation}
    Y(\rho_-=R_1)=e^{i\gamma}K \ Y(\rho_+=cR_1).
    \label{bc field}
\end{equation}

\section{Energy Functional}

Since we are considering a different setup, i.e., two separate spacetimes with a singular timelike shell between them,  it is worth checking the existence of a conserved energy for self-adjoint boundary conditions. In Ref.~\citep{ishibashi}, Ishibashi and Wald defined the energy functional
\begin{equation}
E(\hat{\Psi})=\langle \dot{\hat{\Psi}},\dot{\hat{\Psi}}\rangle+\langle \hat{\Psi},A\hat{\Psi}\rangle, 
\label{energy}
\end{equation}
where $\langle \cdot,\cdot \rangle$ denotes the usual inner product in $L^2(\Sigma,V^{-1}d\Sigma)$. It is easy to check that Eq.~(\ref{energy}) corresponds to the usual (canonical) energy (extracted from the energy-momentum tensor) in globally hyperbolic spacetime. However, it has an extra boundary term (which compensates the exchange of energy-momentum at the boundary).

In our case, we have the total energy of spacetime as the sum of each part:
\begin{equation}
E = E_++E_-,
\end{equation}
where the definition of each energy functional is as follows
\begin{equation}
E_\pm = \frac{1}{2}\int_{\Sigma_t^{\pm}}(\dot{\hat{\Psi}}^2-\hat{\Psi}\vec{\nabla}_\pm^2\hat{\Psi})d\Sigma,
\end{equation}
here we have considered the hypersurface of constant time $\Sigma_t=\Sigma_t^{+}\cup \Sigma_t^{-}$, where $\Sigma_t^{\pm}$ lives in $\mathcal{V}^{\pm}$. In both of these spacetime regions, we use the well-known property below in order to simplify the energy functional
\begin{equation}
    \vec{\nabla}_\pm\cdot\vec{\nabla}_\pm(f^2)=\vec{\nabla}^2_{\pm}(f^2)=2(f\vec{\nabla}_{\pm}^2f+(\vec{\nabla}_{\pm}f)^2),
\end{equation}
 and apply the divergence theorem to obtain the total energy as the sum of a canonical energy functional and a boundary term $ E=E_c+\partial E,$ where  the canonical term for the field energy is
\begin{equation}
E_c =  \frac{1}{2}\int_{\Sigma_t^{-}} (\dot{\hat{\Psi}}_-^2+(\vec{\nabla}\hat{\Psi}_-)^2)+ \frac{1}{2}\int_{\Sigma_t^{+}} (\dot{\hat{\Psi}}_+^2+(\vec{\nabla}\hat{\Psi}_+)^2),
\end{equation}
and the boundary term is
\begin{equation}
\partial E = -\frac{R_1}{2}\int(\hat{\Psi}_-{\partial_{\rho_-}\hat{\Psi}_-}|_{\rho_-=R_1}-\hat{\Psi}_+{\partial_{\rho_+}\hat{\Psi}_+}|_{\rho_+=cR_1})d\phi dz.
\end{equation}

Through the relation \(\vec{\nabla}_{\pm}\cdot(\dot{\hat{\Psi}}_{\pm}\vec{\nabla}_{\pm}{\hat{\Psi}}_\pm)=\dot{\hat{\Psi}}_{\pm}\vec{\nabla}_{\pm}^2\hat{\Psi}_\pm+\vec{\nabla}_{\pm}\hat{\Psi}_\pm\vec{\nabla}_{\pm}\dot{\hat{\Psi}}_\pm\) and the equation of motion \(\ddot{\hat{\Psi}}_\pm=\vec{\nabla}_{\pm}^2\hat{\Psi}_\pm\), we have
\begin{equation}
\frac{dE_c}{dt}=\int R_1(\dot{\hat{\Psi}}_-{\partial_{\rho_-}\hat{\Psi}_-}|_{\rho_-=R_1}-\dot{\hat{\Psi}}_+{\partial_{\rho_+}\hat{\Psi}_+}|_{\rho_+=cR_1})d\phi dz
\end{equation}
so that, adding the temporal derivative of $\partial E$, we get the temporal derivative of the total energy 
\begin{equation}
\begin{aligned}
\dot{E}=\frac{R_1}{2}\int d\phi dz &\Bigg(\left[\dot{\hat{\Psi}}_-{\partial_{\rho_-}\hat{\Psi}_-}|_{\rho_-=R_1}-\hat{\Psi}_-{\partial_{\rho_-}\dot{\hat{\Psi}}_-}|_{\rho_-=R_1}\right]\\&-\left[\dot{\hat{\Psi}}_+{\partial_{\rho_+}\hat{\Psi}_+}|_{\rho_+=cR_1}-\hat{\Psi}_+{\partial_{\rho_+}\dot{\hat{\Psi}}_+}|_{\rho_+=cR_1}\right]\Bigg)
\end{aligned}
\end{equation}
After applying the boundary conditions given by Eq.~(\ref{bc field}), we obtain
\begin{equation}
    \dot{E}=\frac{R_1}{2}\int_{\rho_+=cR_1} d\phi dz (1-\textrm{det}(K))\hat{\Psi}_- \overset{\leftrightarrow}{\partial}_{\rho_-}\dot{\hat{\Psi}}_-,
\end{equation}
and this equation implies that \(\dot{E}=0\) if \(\textrm{det}(K)=1\), which is our criterion for a self-adjoint condition.

\section{Unruh-deWitt Detector}
The Unruh-DeWitt detector consists in a quantum system with two energy levels, $|E_0 \rangle$ and $|E\rangle$, coupled with the scalar field via a monopole interaction. Denote $\Omega = E-E_0$ the energy gap of the detector. When the  detector interacts with the field jumping from the ground state $|E_0 \rangle$ to the excited state $|E \rangle$, we say that the detector  detected a quantum of energy $\Omega$. In the same way, the detector can be deexcited when passing from the excited state $|E \rangle$ to the ground state $|E_0 \rangle$ (in this  $\Omega<0$).

We aim to study the detector's response to excitation and de-excitation using perturbation theory, focusing on its lowest order~\citep{birrell}. Consider the total Hamiltonian of the  system field-detector-interaction given by
\begin{equation}
\mathscr{H} = \mathscr{H}_C + \mathscr{H}_D + \mathscr{H}_I
\end{equation}
where $\mathscr{H}_C$ is the Hamiltonian of the scalar field, $\mathscr{H}_D = \Omega |E \rangle \langle E|$ is the detector's Hamiltonian satisfying $\mathscr{H}_D|E_0\rangle = 0$ and $\mathscr{H}_D|E\rangle = \Omega |E\rangle$, and $\mathscr{H}_I$ is the interaction Hamiltonian. In perturbation theory, we treat $\mathscr{H}_I$ as the perturbative part and $\mathscr{H}_0 = \mathscr{H}_C + \mathscr{H}_D$ as the Hamiltonian which can be solve exactly. The formalism operates on the tensor product of two Hilbert spaces, $\mathcal{H}_D \otimes \mathscr{F}$, with $\mathcal{H}_D$ being the two-dimensional space formed by the detector states $\{|E_0\rangle, |E\rangle\}$, and $\mathscr{F}$ being the Fock space for the scalar field.

Let $\hat{m}(\tau)$ be the monopole moment operator. In the Heisenberg representation:
\begin{equation}
\hat{m}(\tau) = e^{i\mathscr{H}_0 \tau} \hat{m}(0) e^{-i\mathscr{H}_0 \tau}
\end{equation}
By taking $\hat{m}(0) = |E\rangle \langle E_0| + |E_0\rangle \langle E|$, the interaction Hamiltonian (considering the detector as a point-like system) is given by
\begin{equation}
\mathscr{H}_I(\tau) = \lambda \chi(\tau) \left( e^{-i \tau \Omega} |E \rangle \langle E_0| + e^{i \tau \Omega} |E_0 \rangle \langle E| \right) \otimes \hat{\Psi}(x(\tau)),
\end{equation}   
where $\lambda$ is a small coupling constant, $\chi(\tau)$ is a smooth compact support function that `switches' the field-detector interaction on and off. If we assume the initial state $|i\rangle = |E_0\rangle \otimes |0\rangle$ for the system detector+field we have that, after the interaction, the detector has a non-zero probability of being found in its excited state $|E\rangle$, with the field in a state $|\psi \rangle \neq |0\rangle$, giving the final state $|f\rangle = |E\rangle \otimes |\psi\rangle$. For $|\lambda| \ll 1$, the transition amplitude can be calculated in first-order perturbation theory via the Dyson series, for distinct states $i \neq f$
\begin{equation}
\mathcal{A}(\Omega) = -i \lambda \int_{-\infty}^{\infty} \chi(\tau) e^{i \Omega \tau} \langle \psi| \hat{\Psi}(x(\tau)) |0\rangle \, d\tau
\end{equation}
And the transition probability, summing over all possible final states $|1_n\rangle$, is ( see, for instance, Refs. \cite{schlicht} and \cite{satz_2007}) :
\begin{equation}
\mathcal{P}(\Omega) = \lambda^2 | \langle E|\hat{m}(0)|E_0\rangle |^2 \mathscr{F}(\Omega)
\end{equation}
where
\begin{equation}
\mathscr{F}(\Omega) = \int d\tau d\tau' \chi(\tau) \chi(\tau') e^{-i \Omega (\tau - \tau')} G(x(\tau), x(\tau')) \label{funcao resposta}
\end{equation}
The function $\mathscr{F}$ is known as the response function. It is independent of the detector's internal details and is determined by the Wightman Green function $G(x(\tau), x(\tau')) = \langle 0|\hat{\Psi}(x(\tau)) \hat{\Psi}(x(\tau')) |0\rangle$.

 Notice  that the modes of the scalar field when analyzed along the trajectory of the detector at rest, \( x(\tau) = (\tau, \rho_d, \phi_d, z_d) \), are those associated with the inner region \(\mathcal{V}^-\) of the spacetime,  i.e., 
\begin{equation}
{\hat{\Psi}}_{kmq}(x(\tau))=N_{kmq}e^{-i\omega\tau}e^{ikz_d}e^{im\phi_d}J_m(q\rho_d).
\end{equation}
%and when, by symmetry arguments, we fix the coordinates \(z_d = \phi_d = 0\), the field modes take the following form
%\[
%{\hat{\Psi}}_{kmq}(x(\tau)) = N_{kmq}e^{-i\omega\tau}J_m(q\rho_d).
%\]
Keeping the detector at the center of the inner region \(\mathcal{V}^-\) with $\rho_-=\rho_d = 0$, the normalized modes in $\mathcal{V}^-$ take the form\footnote{By symmetry arguments, we fix the coordinates \(z_d = \phi_d = 0\).}
\begin{equation}
{\hat{\Psi}}_{kmq} = 
\begin{cases}
\frac{1}{2\pi\sqrt{2\sqrt{k^2+q^2}}||\psi_{mq}||}e^{-i\omega \tau}, & \text{if } m = 0, \\
0, & \text{if } m \neq 0,
\end{cases}
\end{equation}
where $\|\psi_{mq}\|^2=\int_0^{R_1}|J_m(q\rho')|^2\rho'd\rho'+\int_{cR_1}^{cR_2}|c_1J_{cm}(q\rho')+c_2Y_{cm}(q\rho')|^2\dfrac{\rho'}{c}d\rho'$ is the norm of $\psi_{mq}$ (which is finite, since we are considering a finite value for $R_2$). The response function takes the form
\begin{equation}
\mathscr{F}(R_1, R_2, c, \Omega, \Delta \tau) = \sum_{q} \mathscr{F}_q(R_1, R_2, c, \Omega, \Delta \tau),
\label{truncated}
\end{equation}
with
\begin{equation}
\mathscr{F}_q = \frac{1}{4\pi}\frac{1}{||\psi_{0  q}||^2}\int_{-\infty}^{\infty}dk\frac{|\hat{\chi}(\Omega+\sqrt{k^2+q^2})|^2}{\sqrt{k^2+q^2}},
\end{equation}
where $\hat{\chi}(k)$ is the Fourier transform of $\chi(\tau)$.

Given  the form of the function \(\chi(\tau)\)
\begin{align}
    \chi(\tau) &= \begin{cases}
      \cos^4(\eta \tau), &-\frac{\pi}{2 \eta}\leq \tau\leq\frac{\pi}{2 \eta}\\
       0, &\text{otherwise\,.}
   \end{cases}\label{eq:compact}
\end{align}
as in \citep{cong}, we have
\begin{equation}
\hat{\chi}(y) = \frac{24 \sqrt{\frac{2}{\pi}}\eta^4 \sin(\frac{\pi y}{2\eta})}{y^5 - 20y^3\eta^2 + 64y \eta^4}, \ \text{if} \ \eta > 0,
\end{equation}

\section{Results}
In this section we study the response function of the Unruh-deWitt detector placed at $\rho_-=\rho_d=0$. We analyse its sensitivity to different boundary conditions at $\rho_-=R_1$ ($\rho_+=c R_1$), in addition to the sensitivity to the conicity of the spacetime as considered in Ref.~\citep{cong}. We do not consider the most general case in  Eq.~(\ref{bc field}), but restrict our analysis to a particular class of self-adjoint boundary conditions given by
\begin{equation}
    \gamma = 0 \ \ \text{and} \ \     K = \begin{pmatrix}
        \beta & 0 \\
        0 & 1/\beta
    \end{pmatrix},
\label{matrix simplified}
\end{equation}
where $\beta\in \mathbb{R}$. These boundary conditions are simple enough to simplify many of our calculations but still general enough to exemplify our main considerations related to the freedom on the choice of boundary conditions.

The system of equations specifying the boundary conditions for the radial part of the scalar field is given by
\begin{equation}
\left\{
\begin{array}{l}
    \psi_{mq}(\rho_- = R_1) = \beta \psi_{mq}(\rho_+ = cR_1), \\
    \psi_{mq}'(\rho_- = R_1) = \frac{1}{\beta} \psi_{mq}'(\rho_+ = cR_1), 
\end{array}
\right.\label{condicao_diag}
\end{equation}
along with Dirichlet boundary condition at $\rho_+=cR_2$
\begin{equation}
     \psi_{mq}(\rho_+ = cR_2) = 0.
     \label{Diri}
\end{equation}

Given the form of $\psi_{mq}$ in Eq.~(\ref{bessels}), we can show that Eqs.~(\ref{condicao_diag}) and~(\ref{Diri}) have no solution for $q^2<0$, i.e., there are no bound states. Hence, the eigenvalues of the Sturm-Liouville problem are of the form $\lambda = q^2>0$ and  Eq.~(\ref{condicao_diag}) implies
\begin{equation}
    \left\{\begin{aligned}
         c_1 =&  \frac{1}{2\beta}c\pi \Big[(-\beta^2 qR_1 J_{m-1}(qR_1) + (-1+\beta^2)m J_m(qR_1))Y_{cm}(cqR_1) \\&+  qR_1 J_m(qR_1) Y_{cm-1}(cqR_1)\Big], \\
          c_2 =& \frac{1}{2\beta}c\pi \Big[(\beta^2 qR_1 J_{m-1}(qR_1) - (-1+\beta^2)m J_m(qR_1))J_{cm}(cqR_1) \\&-  qR_1 J_m(qR_1) J_{cm-1}(cqR_1)\Big],
    \end{aligned}\right.
\end{equation}
Notice that the smooth boundary condition is a particular case of Eq.~(\ref{matrix simplified}) with $\beta = 1$. For $\beta \neq 1$, the field and its derivative will be discontinuous. %Moreover, if we multiply the first two lines of equation (\ref{condicao_diag}) side by side, we obtain a conservation relation in $S$ for the function $$F_{mq}(\rho_{\pm}) = \psi_{mq}(\rho_{\pm}) \psi_{mq}'(\rho_{\pm}) = \frac{1}{2} \partial_{\rho_\pm} (\psi_{mq}(\rho_\pm)^2).$$
%In fact,
%\begin{equation}
  %  \psi_{mq}(\rho_-) \psi'_{mq}(\rho_-) |_{\rho_- = R_1} = \psi_{mq}(\rho_+) \psi'_{mq}(\rho_+) |_{\rho_+ = cR_1},
%\end{equation}
%thus, mode-by-mode, the radial flux of the field $\hat{\Psi}$, defined by the function $F_{mq}$, is preserved when crossing the hypersurface $S_1$ joining the regions $\mathcal{V}^-$ and $\mathcal{V}^+$under diagonal boundary conditions.
Besides, notice that $c_i \rightarrow - c_i$ when $\beta \rightarrow -\beta$ for $i = 1, 2$, then we consider only $\beta>0$ in our investigations. The values of $q$ satisfying Dirichlet boundary condition \(\psi_{mq}(q c R_2)=0\) can be found numerically and depend crucially on the choice of the boundary condition $\beta$. We truncate the sum in Eq.~(\ref{truncated}) by considering $N$ terms such that $|\mathscr{F}_{q_{N+1}}|<10^{-7}$.

\begin{figure}[htb!]
    \centering
    \includegraphics[width=\linewidth]{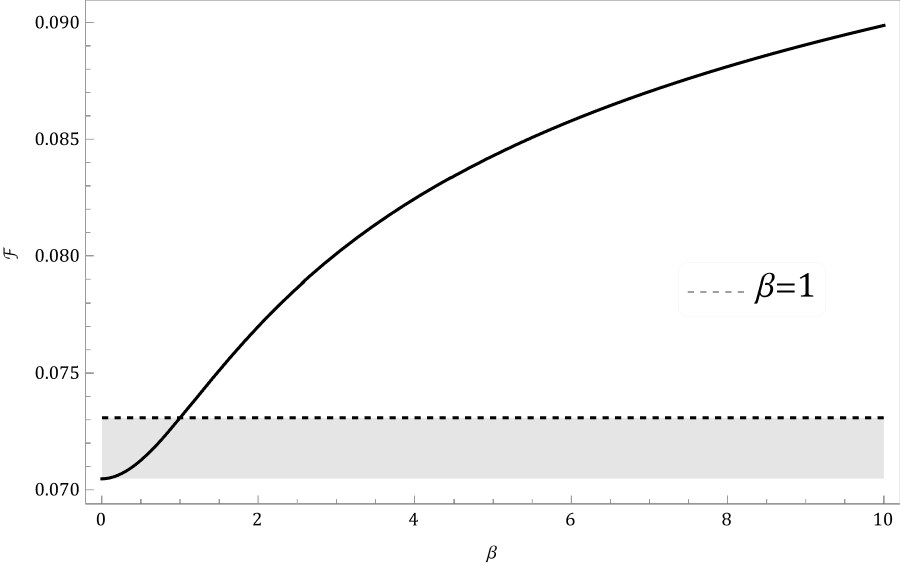}
    \caption{\small The figure shows the graph of \(\mathscr{F}_c\) as a function of \(\beta\) in the interval \(0 \leq \beta \leq 10\). Here, the parameters are \(c=2\), \(R_1=\Delta\tau=1\), \(R_2=5\), \(\rho_d=0\), and \(\Omega=1/2\). The dashed line represents \(\mathscr{F}_{c=2}\) with \(\beta=1\).}
    \label{fig1}
\end{figure}

In Fig.~\ref{fig1}, we consider $R_1=\Delta\tau=1$ and $c=2$. In this scenario , the detector is classically isolated from the singular shell at $\rho_-=R_1$ ($\rho_+=c R_1$). We observe that  the response function \(\mathscr{F}\) grows  with $\beta$ and is finite in the limit $\beta\rightarrow 0$. This shows that the response function for the smooth condition $\beta=1$ found in Ref.~\citep{cong} is (only) a particular case corresponding to one specific choice\footnote{Since the spacetime is singular at the shell connecting $\mathcal{V}^-$ and $\mathcal{V}^+$, the theory of general relativity  does not give a precise prescription for  propagating fields. Hence,  there is no preferred boundary condition and it is impossible to measure any effect of the conicity on the outer region without considering the (arbitrary) choice of the boundary condition at the connecting surface.} for the boundary condition $\beta$.

\begin{figure}[htb!]
    \centering
    \includegraphics[width=\linewidth]{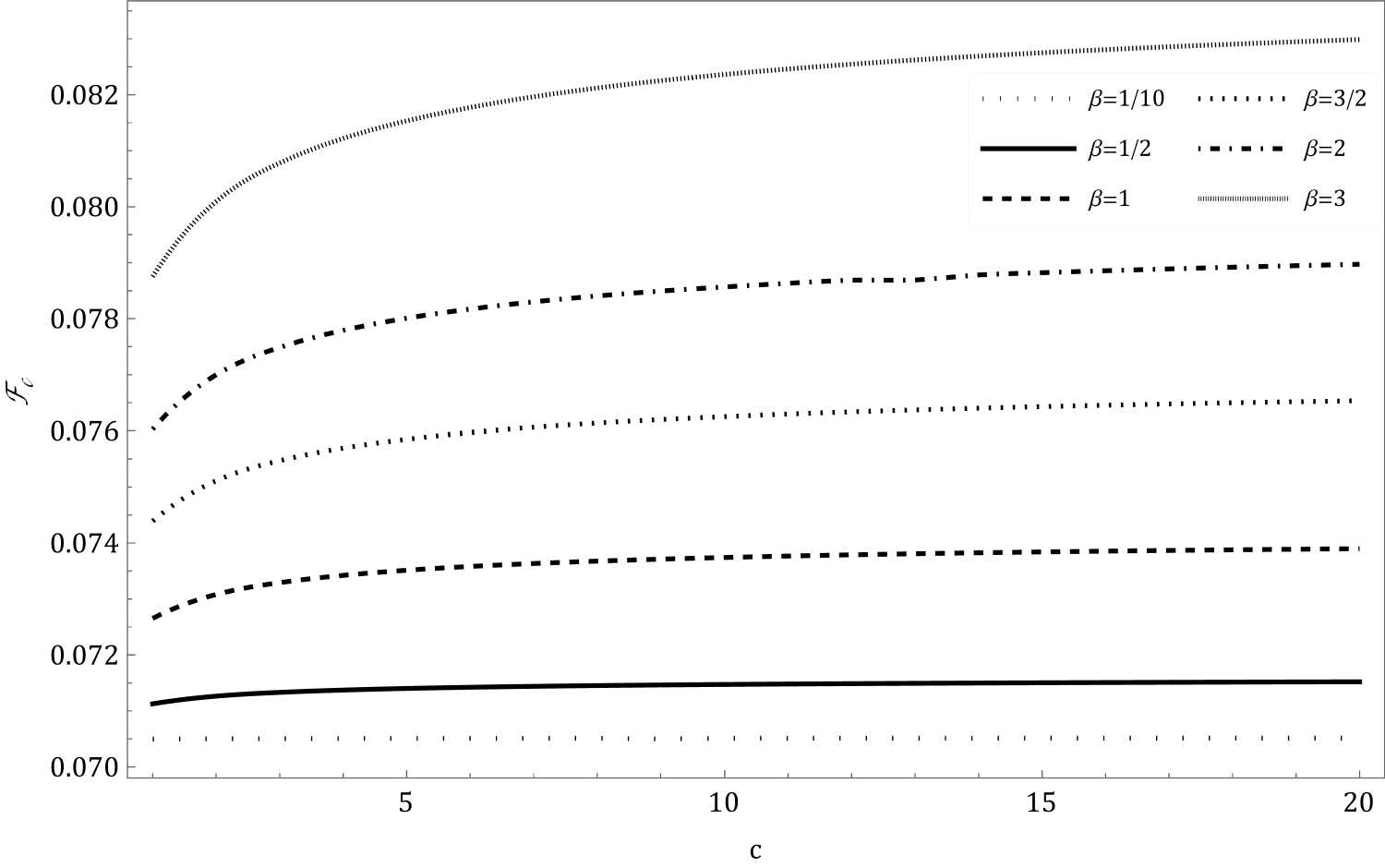}
    \caption{\small The figure presents the curve of \(\mathscr{F}_c\) as a function of \(c\) for different values of \(\beta\). Here, the parameters considered are \(R_1=\Delta\tau=1\), \(R_2=5\), \(\rho_d=0\), and \(\Omega=1/2\). }
    \label{fig2}
\end{figure}

Fig.~\ref{fig2} shows the dependence of the response function on the conicity for several values of the boundary condition $\beta$. The plots indicate that, when only quantum effects take place,  the detector's sensitivity to the conicity changes significantly depending on the choice of the boundary condition. In particular, when $\beta\to 0$, the detector \sout{looses}  {appears to reduce} sensibility with respect to the conicity. { As illustrated in Figure~\ref{fig2}, the detector becomes increasingly insensitive to variations in the conicity parameter within the range \( c \in [0, 20] \). In particular, the response function curves progressively flatten as \( \beta \) decreases, indicating that the excitation probability becomes nearly independent of conicity.
}

\begin{figure}[htb!]
    \centering
    \includegraphics[width=\linewidth]{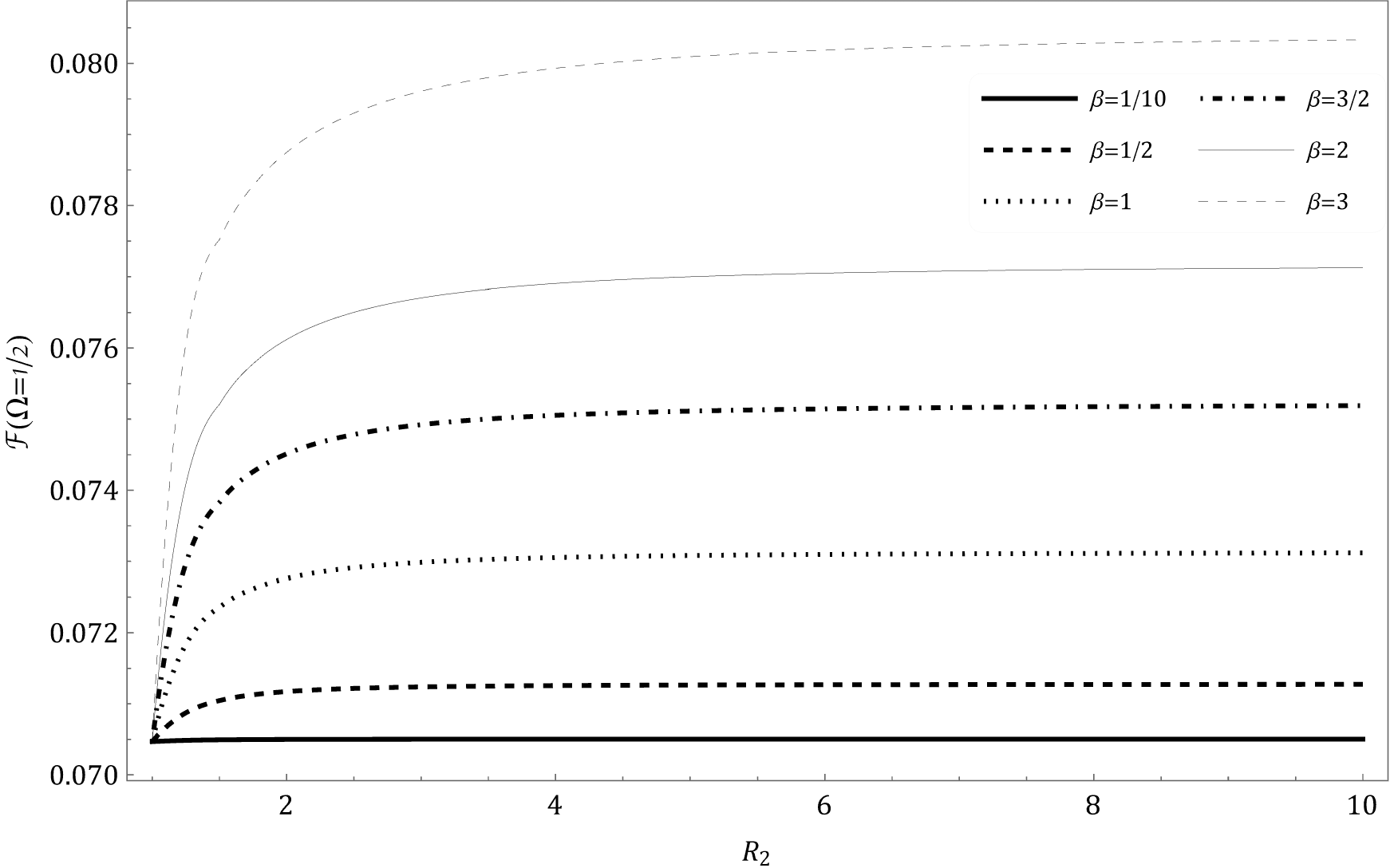}
    \caption{\small The graph shows the response function as a function of \(R_2\) with several diagonal boundary conditions. The parameters used are \(c=2\), \(\rho_d=0\), \(R_2=5\), \(\Omega=1/2\), and \(\Delta\tau=1\).}
    \label{fig6}
\end{figure}

Finally, Fig.~\ref{fig6} shows that,  when $R_2\to\infty$, i.e., when we approximate an asymptotically conical spacetime, the response function tends to a constant depending crucially on the boundary condition. 

\begin{figure}[ht!]
    \centering
    \includegraphics[width=\linewidth]{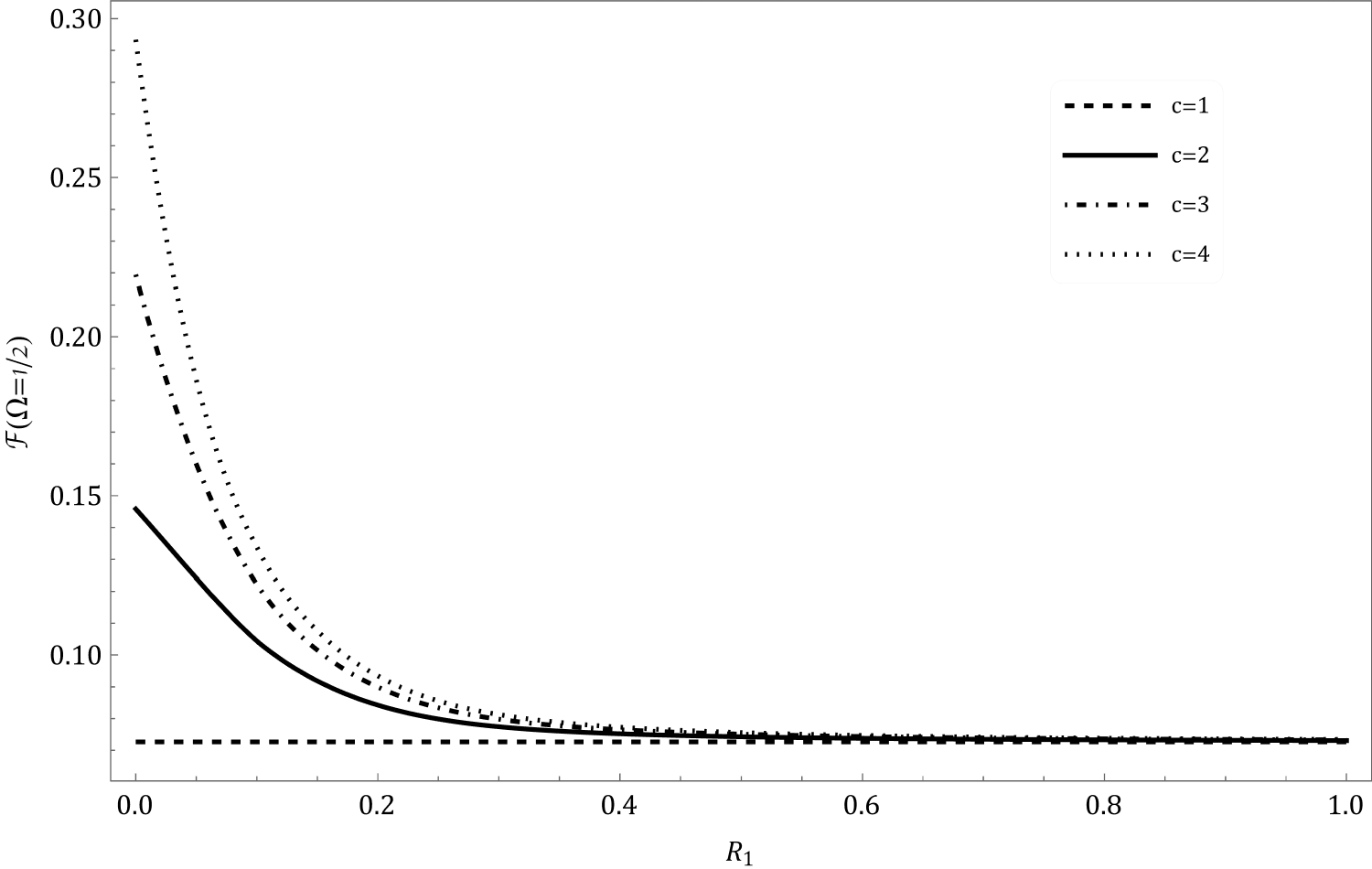}
    \caption{\small The graph shows the dependence of the response function on the radius \(R_1\) for conicity values \(c=1, 2, 3\), and 4 and for $\beta=1$. Here, the parameters used are \(\rho_d=0\), \(R_2=5\), \(\Omega=1/2\), and \(\Delta\tau=1\).}
    \label{fig3}
\end{figure}

\begin{figure}[h!]
    \centering
    \includegraphics[width=\linewidth]{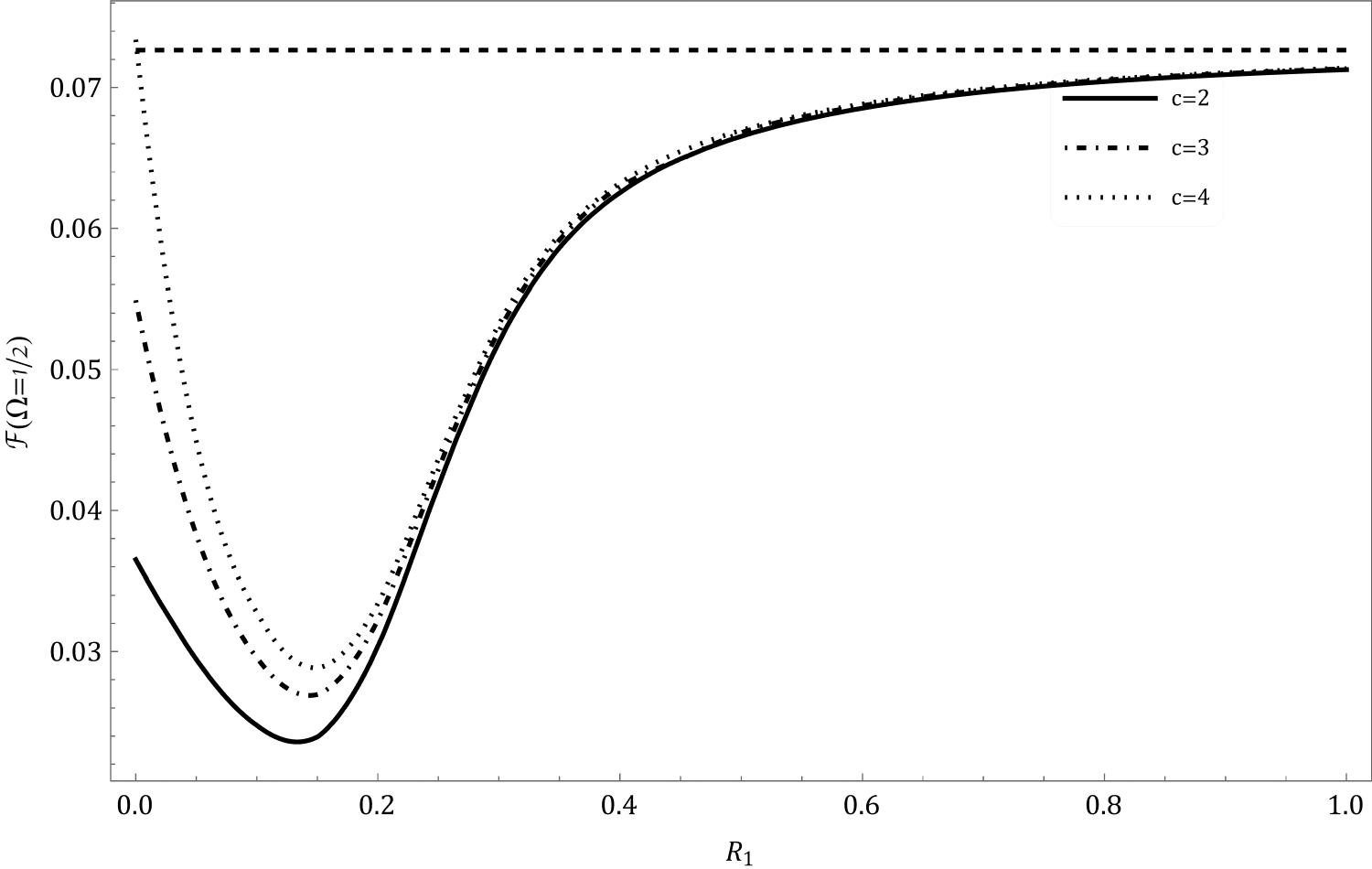}
    \caption{\small The graph shows the response function as a function of \(R_1\) for the diagonal boundary condition with \(\beta=1/2\) and conicity \(c= 2, 3,\) and \(4\). The dashed curve represents the response function with  \(\beta=c=1\). The parameters used are \(\rho_d=0\), \(R_2=5\), \(\Omega=1/2\), and \(\Delta\tau=1\).}
    \label{fig4}
\end{figure}

\begin{figure}[h!]
    \centering
    \includegraphics[width=\linewidth]{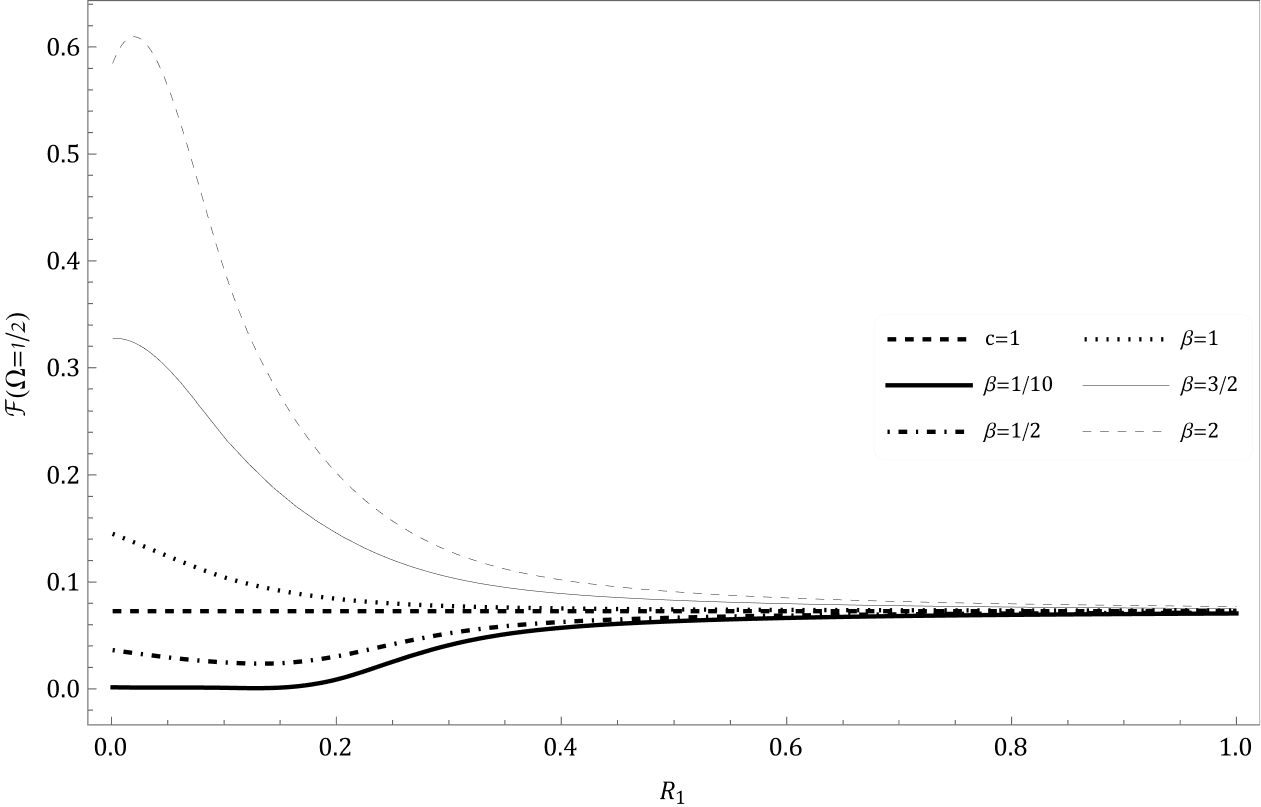}
    \caption{\small The graph shows the response function as a function of \(R_1\) for the conicity $c=2$ and diagonal boundary condition $\beta=1/10,1/2,1$ and $2$. However, the dashed curve represents the response function under smooth boundary condition, $\beta=1$ and defict-free, $c=1$. The parameters used are \(\rho_d=0\), \(R_2=5\), \(\Omega=1/2\), and \(\Delta\tau=1\).}
    \label{fig5}
\end{figure}

 We can also study the sensitivity of the detector to both the conicity and boundary condition when classical effects becomes relevant.  By fixing the interaction time $\Delta \tau=1$ and varying the inner radius $R_1$, we can see what happens when the detector also interacts  classically with the shell. Figs~\ref{fig3},~\ref{fig4} and~\ref{fig5}  show that the graphics {begin to split smoothly around} $R_1=0.5$,  when a signal of light has time to leave the detector, interact with the shell and go back to the detector. We can also see that the split is much more evident when we fix the conicity  and vary the boundary condition.


{Around \( R_1 = 0.5 \), the separation between the response functions corresponding to different parameter values emerges smoothly and continuously. It is worth emphasizing that these response functions remain distinct even for \( R_1 \gg 0.5 \); however, in this regime, the magnitude of the differences becomes significantly smaller. This behavior reflects the fact that, beyond \( R_1 = 0.5 \), the detector can probe the conical geometry only through quantum effects, and the resulting variations in the response function are therefore more subtle.}

\section{Conclusions}
In summary, we studied the Unruh-DeWitt detector's  sensitivity to boundary conditions, even when their classical  effects are absent, i.e., when the surface where these conditions are applied lies beyond the detector's light-crossing region. We considered the same framework studied in Ref.~\citep{cong}, with a detector placed on a flat inner region connected to a conical outer region through a singular shell and boundary condition turns out to be necessary at this singular timelike surface. In Ref.~\citep{cong}, the field and its derivative were supposed to be continuous at the singular shell. However, given that there is no definite prescription for fields at singular shells (general relativity is not valid at singularities), continuity of fields (and its derivative) corresponds to only one possible choice of boundary conditions at singular surfaces. Hence, the resulting response function measures not only topological properties of the spacetime but also an arbitrary condition at the singular shell. Finally, we highlight that our results can be extrapolated to more general spacetimes connected by a singular shell.

\section*{Acknowledgements}
J. M. S.  thanks the Coordination for the Improvement of Higher Education Personnel - Brazil (CAPES) - Finance Code 001 (process number 88887.674194/2022-00). J. P. M. P.
thanks the support provided in part by Conselho
Nacional de Desenvolvimento Científico e Tecnológico
(CNPq, Brazil), Grant No. 311443/2021-4, and Fundação
de Amparo \`a Pesquisa do Estado de São Paulo (FAPESP)
Grant No. 2022/07958-4.

%I am grateful to CAPES and to the Graduate Program in Applied Mathematics at the Institute of Mathematics, Statistics, and Scientific Computing (IMECC) of the University of Campinas (UNICAMP). I extend my sincere thanks to the faculty for their guidance and dedication. Additionally, I express my profound gratitude to the program staff for the exceptional structure and support provided throughout the development of this project.



\begin{thebibliography}{99} 

\bibitem[Wald(1984)]{wald}
Wald, R. M.,  {\it General Relativity}, Chicago University Press, Chicago 1984.

\bibitem[Wald(1980)]{wald_1980}
Wald, R. M.,  {\it Dynamics in nonglobally hyperbolic, static space-times}, J. Math. Phys. {\bf 21}, 2802 (1980).

\bibitem[Ishibashi et al.(2003)]{ishibashi}
Ishibashi, A. and Wald, R. M.,  {\it Dynamics in non-globally-hyperbolic static spacetimes: II. General analysis of prescriptions for dynamics}. Class. Quant. Grav. {\bf 20}, 3816 (2003).


\bibitem[Everitt et al.(1986)]{everitt}
Everitt, W. N. and Zettl, A., {\it Sturm-Liouville differential operators in direct sum spaces},  Rocky Mountain J. Math. {\bf 16},  497 (1986).

\bibitem[Zettl(2005)]{zettl}
Zettl, A. {\it Sturm-Liouville theory}, American Mathematical Society, 2005.


\bibitem[Wang et al.(2007)]{wang}
Wang, A.,   Sun, J. and Zettl, A., {\it Two interval Sturm–Liouville operators in modified Hilbert spaces},  J. Math. Anal. Appl. {\bf 328}, 390 (2007).

\bibitem[Pitelli et. al.(2024)]{pitelliwire}
J. P. M. Pitelli, R. A. Mosna and  F. F. Souto, {\it Quantum mechanics on sharply bent wires via two-interval Sturm-Liouville theory}, J. Math. Phys. {\bf 65}, 062104 (2024).

\bibitem[Andrews(1981)]{andrews}
M. Andrews, {\it Matching conditions on wave functions at discontinuities of the potential}, Am. J. Phys. {\bf 49}, 281–282 (1981).



\bibitem[Cong et al.(2021)]{cong}
Cong, W,  Bi\v c\'ak, J., Kubiz\v n \' ak, D. and Mann., R. B.,  {\it Quantum detection of conicity},  Phys. Lett. B {\bf 820}, 136482 (2021).

\bibitem[Reed et al.(1975)]{simon}
Reed, M. and Simon, B., {\it Methods of Modern Mathematical Physics II: Fourier Analysis, Self-Adjointness}, Academic Press, San Diego, 1975.

\bibitem[Birrell et al.(1982)]{birrell}
Birrell, N. D. and Davies, P. C. W.,  {\it Quantum Fields in Curved Space}. Cambridge University Press, Cambridge, 1982.


\bibitem[Schlicht(2004)]{schlicht} S. Schlicht, {\it Considerations on the Unruh effect: causality and regularization}, Class.
Quantum Grav. {\bf 21}, 4647 (2004).

\bibitem[Satz(2007)]{satz_2007} Satz,  A. {\it Then again, how often does the Unruh–DeWitt detector click if we switch it carefully?}, Classical and Quantum Gravity {\bf 24}, 1719–1731 (2007).

\end{thebibliography}
\end{document}